\documentclass[12pt,preprint]{aastex}

\shorttitle{Ultraviolet Spectroscopy of Solar-Mass Stars}
\shortauthors{Jeffrey L. Linsky et al.}
\slugcomment{Draft \today}

\begin{document}

\newcommand{\php}[0]{\phantom{--}}
\newcommand{\kms}[0]{km~s$^{-1}$}
\newcommand{\fluxunits}{[~erg~cm$^{-2}$~s$^{-1}$~\AA$^{-1}$]}

\title{ULTRAVIOLET SPECTROSCOPY OF RAPIDLY ROTATING SOLAR-MASS STARS: 
EMISSION-LINE REDSHIFTS AS A TEST OF THE SOLAR-STELLAR CONNECTION}

\author{Jeffrey L. Linsky\altaffilmark{1}}
\affil{JILA, University of Colorado and NIST, 440UCB Boulder, CO 80309-0440,
USA}
\email{jlinsky@jilau1.colorado.edu}

\author{Rachel Bushinsky}
\affil{APS, University of Colorado, 391UCB Boulder, CO 80309-0391, USA}

\author{Tom Ayres\altaffilmark{1}}
\affil{CASA, University of Colorado, 593UCB Boulder, CO 80309-0593, USA}

\and

\author{Kevin France}
\affil{CASA, University of Colorado, 593UCB Boulder, CO 80309-0593, USA}

\altaffiltext{1}{Guest Observer, NASA/ESA {\it Hubble Space Telescope} and 
User of the Data Archive at the Space Telescope 
Science Institute. STScI is operated by the Association of Universities for 
Research in Astronomy, Inc., under NASA contract NAS 5-26555. These 
observations are associated with programs \#11532, \#11534, and \#11687.}

\begin{abstract}

\noindent We compare high-resolution ultraviolet spectra of the Sun 
and thirteen solar-mass main sequence stars with different rotational periods 
that serve as proxies for 
their different ages and magnetic field structures. In this the second 
paper in the series, we study the dependence of ultraviolet emission-line
centroid velocities on stellar rotation period,
as rotation rates decrease from that of the Pleiades star HII314 
($P_{\rm rot}=1.47$ days) to $\alpha$~Cen~A ($P_{\rm rot}=28$ days). 
Our stellar sample of F9~V to G5~V stars consists of six stars observed 
with the Cosmic Origins Spectrograph
on HST and eight stars observed with the Space Telescope Imaging Spectrograph
on HST. We find a systematic trend of increasing redshift with more rapid 
rotation (decreasing rotation period) that is similar to the increase in line
red shift between quiet and plage regions on the Sun.
The fastest-rotating solar-mass star in our study, HII314,
shows significantly enhanced redshifts at all temperatures above $\log T=4.6$, 
including the corona, which is very different from the redshift pattern 
observed in the more slowly-rotating stars. This difference in the redshift 
pattern suggests that a 
qualitative change in the magnetic-heating process occurs near 
$P_{\rm rot}=2$ days. We propose that HII314 is an example of a solar-mass star
with a magnetic heating rate too large
for the physical processes responsible for the redshift pattern to operate 
in the same way as for the more slowly rotating stars. HII314
may therefore lie above the high activity end of the set of 
solar-like phenomena that is often called the ``solar-stellar connection''. 

\end{abstract}

\keywords{stars: chromospheres --- stars: individual (HII314, EK Dra, 
$\zeta$ Dor, $\pi^1$ UMa, $\chi^1$ Ori, HD 165185, HD 25825, HD 97334,
$\kappa^1$ Cet, HD 73350, HD 59967, HD 209458, Sun, $\alpha$ Cen A) ---  
stars:solar-type --- Sun:chromosphere --- Sun:UV radiation}

\section{INTRODUCTION}

After their pre-main sequence evolution, solar-mass stars reach the zero-age 
main sequence  as rapid rotators with strong magnetic fields. 
As they age on the main sequence, the torque of their magnetized winds slows
their rotation rate \citep[e.g.,][]{Matt2008}, producing a wide range of 
changes in their magnetic structure 
and emission properties. \citet{Wilson1963} and \citet{Kraft1967},
among others, showed that the strength of the chromospheric Ca~II lines 
decreases with age and rotational velocity for solar-type stars. 
Other changes also occur in the X-ray and
radio emission, starspot properties, and magnetic field structure. 
In this series of papers, we explore the evolution of the ultraviolet spectra 
of solar-mass stars as they slow down with age on the main sequence.

In a series of four papers, we measure and interpret the properties of 
ultraviolet emission lines and continua formed above the photospheres of a 
sample of solar-mass dwarf stars that differ primarily in the rotation rates 
that govern their magnetic activity. 
The emission lines are formed in the chromospheres
and higher atmospheric layers that are heated by magnetic processes that 
weaken as slower rotation decreases the dynamo generation of 
magnetic fields in these convective stars. The properties of the emission lines
provide qualitative measures of the magnetic heating rates. 
This stellar intercomparison, using rotation period as the independent 
variable, is now feasible because of the availability of
high-resolution spectra of 13 solar-mass stars observed with the Cosmic 
Origins Spectrograph (COS) and Space Telescope Imaging Spectrograph (STIS)
on the Hubble Space Telescope (HST). Our intercomparison of solar-mass stars 
follows on the ``Sleuthing the Dynamo'' study of young cluster stars 
\citep{Ayres1996,Ayres1999}, which was based on lower-resolution UV spectra 
obtained with first generation instruments on HST.
Although the age and rotation dependence
of the ultraviolet and X-ray emission of solar-mass stars, often referred to 
as  the ``Sun in Time'' project, is an active field of research 
\citep[e.g.,][]{Ribas2005, Pizzolato2003,Simon1985}, 
we provide a new and more detailed perspective using high-resolution spectra. 

In the first paper of the series \citep{Linsky2012}, we measured the 
continuum flux between the emission lines in the 1150--1500~\AA\ region 
for seven solar-mass stars and the Sun.
We found that the continuum brightness temperatures ($T_B$) 
increase systematically with increasing rotation rate: the $T_B$ of the
slowest rotator ($\alpha$~Cen~A) is similar to that of solar regions with 
very weak magnetic fields, and the $T_B$ of the fastest rotators 
(HII314 and EK Dra) are
similar to solar regions with strong magnetic fields (bright plages). 
Thus the $T_B$ range seen in solar-mass stars 
with rotation periods ($P_{\rm rot}$) between 1.47 and 28 days corresponds 
to the range in $T_B$ seen from lowest to highest magnetic-field regions on 
the Sun. This result provides a beautiful example of the solar-stellar 
connection.

In the next three papers in the series, we explore three other aspects of the
solar-stellar connection. In the present paper, 
we measure the wavelength shifts of
ultraviolet emission lines in a sample of 14 solar-mass stars including 
the Sun to study the dependence of redshift on rotation period. In the next 
two papers, we will study the line-flux saturation and line-width dependencies
on rotation period. A major question that we will address is whether these
four indicators of stellar magnetic activity are consistent in that they show
roughly the same dependence on stellar rotation velocity.

The decrease in ultraviolet emission-line, X-ray, and EUV fluxes as stars age
on the main sequence has often been characterized by power laws. For example,
\citet{Skumanich1972} proposed that the chromospheric Ca~II line flux and
the rotation velocity both decay with time as $t^{-1/2}$.
In their analysis of the early IUE observations,  
\citet{Ayres1981} showed that the fluxes of chromospheric and 
transition region lines and the X-ray fluxes of spectral type F--K dwarf stars 
are correlated with different power law slopes, indicating that 
age-dependent heating occurs
in the various atmospheric layers but at different rates and perhaps by 
different processes. Subsequent studies
of solar-type main sequence stars [e.g., \citet{Ayres1996,Cardini2007}]
have provided more
accurate slopes to the power-law correlations by using more extensive and 
higher quality UV and X-ray data sets. In their comprehensive study of the
1--1700~\AA\ irradiance of 6 solar-mass stars and the Sun covering ages
$(0.5-1.0)\times 10^8$ yr (EK Dra) to $6.7\times 10^9$ yr ($\beta$ Hyi), 
\citet{Ribas2005} 
obtained power law relations, $F=at^{-b}$, between radiative flux in 
different wavelength bands and stellar age ($t$). For the integrated flux of
UV emission lines, the power-law index increases from about $b=0.70$ for 
low chromosphere lines (C~I, H~I, and O~I) to about $b=1.0$ 
for hot transition region lines (Si~IV, C~IV, and O~VI). 
Thus the young rapidly-rotating Sun at age $10^8$ yr likely had 
fluxes relative to the present Sun enhanced by a factor of 30 for chromospheric
emission lines to 1000 times larger for X-rays. 

There is empirical evidence that magnetic-field structures are qualitatively 
different in very active solar-mass stars compared to less active stars. 
While the Sun never shows sunspots at the pole or at very high latitudes, 
$\xi$ Boo A (G8~V), the solar-like star just to the right of the break in
the mass flux-X-ray correlation \citep{Wood2005}, shows high-latitude spots 
\citep{Toner1988} and Zeeman-Doppler images (ZDI) show an inclined dipole 
and a large-scale toroidal magnetic field \citep{Petit2005} that is 
very different from the Sun. This pattern
of a dipolar magnetic field with a torroidal component is detected in
ZDIs of other rapidly rotating solar-mass stars \citep{Donati2009}. 
Simulations by \citet{Schrijver2001} of the magnetic field in  
very active Sun-like stars that differ from the Sun only by the imposed 
rate of magnetic flux emerging into the photosphere from below 
show a major change from the
solar magnetic-field pattern. When the imposed rate of magnetic-field 
emergence is 30 times that of the Sun, a strong magnetic-field polar cap 
forms surrounded by a torroidal ring of opposite polarity.
The polar cap likely includes large starspot clusters. 
The magnetic flux on the 
surface of this very active Sun-like star is similar to that of a solar-mass 
stars with $P_{rot}=6$ days. The rotation period of $\xi$ Boo A is 6.2 days
\citep{Noyes1984}.  

The observations just summarized point to a qualitative difference between the
magnetic-field structure of rapidly rotating solar-mass stars 
compared to slowly rotating stars like the Sun; 
the magnetic field simulations confirm this difference. Given the 
qualitative difference in magnetic-field structure, one might expect
that the kinematics of warm plasma in the atmosphere 
layers between the photosphere and corona would also be very different, 
because a major fraction of the
energy from magnetic-field reconnection events is converted into heat that is 
then radiated from these atmosphere layers.
Until recently, only a few F- and G-type stars were bright enough to be 
studied with high-resolution UV spectroscopy -- $\alpha$~Cen~A 
\citep{Wood1997,Pagano2004}, Procyon \citep{Wood1997},
$\zeta$~Dor \citep{Redfield2001}, and EK Dra \citep{Ayres-France2010}. 
These studies
raised many of the questions that form the basis for the present systematic
study of stars very close in spectral type and mass to the Sun. 
The very efficient COS instrument 
on HST has increased the number solar-mass stars available for this study,
in particular, the more distant rapidly rotating stars like HII314
located in the Pleiades cluster at 134 pc.

In Section 2 we describe the observations obtained with COS and STIS 
that form our data base. 
Section 3 describes our procedures for extracting emission-line redshifts,
and Section 4 compares the stellar and solar redshifts. 
In Section 5, we summarize our results.

\section{DATA SET AND HST OBSERVATIONS}

We have searched the MAST\footnote{http://archive.stsci.edu} 
archive and the StarCAT STIS spectral catalog \citep{Ayres2010}
for UV spectra (1150--1600~\AA) of solar-mass stars, which for this study
are defined as those stars with spectral types between F9~V and G5~V. 
These spectral types correspond to the mass range 0.92--1.1 solar masses. 
We identify 13 stars observed with the COS G130M and STIS E140M and E140H 
gratings that, including the Sun, we call 
the high-resolution group (see Tables 1 and 2). One star, $\chi^1$~Ori,
was observed by both COS and STIS. The spectral resolving
powers $R = \lambda/\Delta\lambda = 17,000$ (COS G130M), 45,000 (STIS E140M), 
and 114,000 (STIS E140H) allow us to analyze line-profile shapes and 
Doppler shifts. Tables 1 and 2
provide information on the observing programs, stellar ages, cluster 
memberships, rotation periods ($P_{\rm rot}$), rotation velocities ($v$sini), 
and spectral types of stars in both groups.

Our collection of solar-mass stars includes the stars observed with IUE in
the ``Sun in Time'' program of \citet{Ribas2005} and the young stars observed 
by \citet{Ayres1996} and \citet{ Ayres1999} with the GHRS 
G140L and FOS G130H gratings.  
We include only those stars with usable S/N in the observed UV emission lines.

The Pleiades Cluster 
G1~V star HII314 was selected for COS observations because it has the highest
UV line fluxes of a solar mass star in the \citet{Ayres1999} sample. 
We adopt the rotational period $P_{\rm rot}=1.47$ days obtained by 
\citet{Rice2001} in their Doppler-imaging study of the star. EK Dra
[$P_{\rm rot}=2.6050\pm 0.0003$ days, \citet{Strassmeier1998}]
is a member of the Pleiades moving group. $\pi^1$~UMa [$P_{\rm rot}=4.89$
days, \citet{Gaidos2000}], $\chi^1$~Ori [$P_{\rm rot}=5.104$ days,
\citet{King2005}], and HD~165185 [$P_{\rm rot}=5.90$ days, \citet{Saar1997}] 
are members of the Ursa Major moving group. HD~25825 is a member of the Hyades 
cluster, and $\zeta$~Dor [$v$sini=14.8~\kms\ corresponding to
$P_{\rm rot}\leq4.0$ days if $R_{\star}/R_{\odot}=1.15$) 
\citet{Schroeder2009}],
$\kappa^1$~Cet [$P_{\rm rot}=9.24$ days, \citet{Barnes2007}],
HD~97334 [$P_{\rm rot}=7.60$ days, \citet{Hempelmann1995}], HD~73350, HD~59967,
HD~209458, and $\alpha$~Cen~A 
[$P_{\rm rot}=28\pm 3$ days, Barnes 2007] are field stars. 
The rotational period of 11.4 days for HD~209458 was measured by 
\citet{Silva-Valio2008} from occultations of starspots by its transiting 
planet. We estimate that the rotational period for HD~25825 is 
$P_{\rm rot}\approx 6.5$ days based on the measured periods of Hyades G0~V 
stars \citep{Radick1987}, and the mean period for stars with B--V=0.60 and
Hyades age \citep{Barnes2010}.
Rotational periods for the other stars were obtained from
\citet{Pizzolato2003}. 

The HST spectra were obtained from several programs. HII314 was observed by COS
in Program 11532 (J. Green PI).  EK Dra, $\pi^1$~UMa, $\chi^1$~Ori and
HD~25825 were observed as part of the Fe~XXI emission line SNAP 
program 11687 (T. Ayres PI). HD~209458 was observed in program 11534 
(J. Green PI) by \citet{Linsky2010} to study its transiting planet. 
The STIS E140H spectrum of $\alpha$~Cen~A
is published by \citet{Pagano2004}, and the STIS E140M spectra of 
$\chi^1$~Ori and $\kappa$~Cet were obtained from the HST archive. 
Table~3 is the observing log for these COS and STIS observations.

\subsection{COS Observations}

A description of the COS instrument and on-orbit performance  
can be found in \citet{Green2012} and \citet{Osterman2011}.
We observed HII314 (V1038 Tau) with the COS G130M grating on 2009 December 16 
for a total of 3209 seconds. We requested observations for four
orbits, but on the last orbit HST failed to reacquire the target on time, 
leading to a blank spectrum. We observed HII314
with three central wavelength settings ($\lambda$1300,
$\lambda$1309, and $\lambda$1318) to minimize fixed-pattern noise in the
detector. The resulting spectral range was $1134\leq \lambda \leq 1459$, 
with a resolving power of $R\approx$ 17,000--18,000. On 2010, October 11 
we reobserved HII314 using the G160M grating. 
 
We retrieved the calibrated COS data files from
MAST\footnote{http://archive.stsci.edu} and cross-correlated the individual
data files obtained for the different grating settings to determine a common 
wavelength scale using the 
COADD\_x1d.pro software package written in IDL by Charles Danforth.
We found it necessary to reprocess the far-UV observations with this 
custom version of  
CALCOS\footnote{We refer the reader to the cycle 18 COS Instrument Handbook 
for more details: 
http://www.stsci.edu/hst/cos/documents/handbooks/current/cos\_cover.html}
v2.11, because incomplete pulse-height screening produced residual
spurious features in the coadded spectra.
The COS G130M spectrum is presented in Figure~1 together with other COS and 
STIS spectra.

The SNAP program consisted of single orbit observations using only the COS G130
grating at the $\lambda$1291 central wavelength setting that covered the 
wavelength region 1290--1430~\AA. For most of the SNAP
observations, the 1134--1274~\AA\ region of the detector was turned off to 
avoid overexposure in the Lyman-$\alpha$ line.
Exposure times for the four stars in the SNAP program selected for this study 
were between 1160 and 1300 seconds. We reduced these data in the same way 
as described for HII314, except that there was only one wavelength setting,
although FP-POS positions 3 and 4 were utilized.

We also include the combined out-of-transit observations of 
HD~209458 consisting of 20,694 seconds of exposure at both orbital quadratures 
and secondary eclipse of HD~209458b. These observations are described by 
\citet{France2010b} and \citet{Linsky2010}.

\subsection{STIS Observations}

We include in this study STIS moderate resolution E140M spectra of the field
stars $\zeta$~Dor, HD~97334, HD~73350, HD~59967, and $\kappa^1$~Cet and 
the Ursa Major moving group stars HD~165185 and $\chi^1$~Ori, which was
also observed with COS. The STIS observations were obtained with
program 8280 (PI T. Ayres) using the 0.2 x 0.2 aperture. 
The spectra cover the range 1140--1729~\AA\ 
with a spectral resolution of R$\approx$~40,000. 
$\kappa^1$~Cet was observed for 7805 seconds on September 19, 2000, and
$\chi^1$~Ori was observed by STIS for 6770 seconds on October 3, 2000.
We retrieved the STIS spectra from 
StarCAT\footnote{http://casa.colorado.edu/$\sim$ayres/StarCAT/} 
provided by Thomas Ayres. These spectra have well-calibrated flux and 
wavelength scales and did not require further processing.
Figure~1 shows spectra of ten solar-mass stars observed by COS and STIS. 
Included for 
comparison is the STIS E140H spectrum of the slowly rotating near solar twin 
$\alpha$~Cen~A \citep{Pagano2004}. Although the spectra appear very similar, 
the flux scales increase with more rapid rotation. We will discuss this
increase in emission and the resulting saturation of line fluxes in a 
subsequent paper using these observations.

$\alpha$~Cen~A was recently reobserved by STIS using the E140M and  
E140H gratings. Results from this observing program led by 
T. Ayres will be presented elsewhere, but we include here measurements of the 
radial velocities of the Si~IV and C~IV lines.

\subsection{Comparison Solar Irradiance Measurements}

We compare the stellar observations of solar mass stars with solar
irradiance measurements, which are flux values of 
the Sun viewed as a distant point source like our stellar flux measurements.
We use the solar-irradiance reference spectra obtained with the Solar Radiation
and Climate Experiment (SORCE) on the Solar-Stellar Irradiance Comparison 
Experiment II (SOLSTICE II) \citep{Woods2009, McClintock2005, Snow2005}.
These data cover the 1150--3200~\AA\ spectral range with 1.0~\AA\ resolution. 
We selected this data set because the absolute flux calibration is accurate to
about 5\% in the UV and is cross-checked against B and A-type stars. Scattered
and stray light are removed from these data. We use here the March 25--29, 
2008 data set that represents the Sun close to minimum with a Zurich-sunspot
number of 2 and an average 10.7-cm radio flux of $69\times 10^{-22}$ 
Wm$^{-2}$Hz$^{-1}$.

Although the Sun is $3 10^{12}$ times brighter than, for example, 
$\chi^1$~Ori, the relative number of photons observeable by SORCE and COS 
is only $10^4$ considering the 
different aperture sizes (0.1x0.1 nm for SORCE vs 2.4 m diameter for HST),
observing times (180 seconds for SORCE vs 1300 seconds for COS),
and resolution elements (0.1 nm for SORCE and 0.0076 nm for COS). If the two
instruments have comparable efficiences, then the S/N of the SORCE FUV spectra 
should be 30 times that of the COS FUV spectra binned to the SORCE resolution.

\subsection{Extraction of the Emission-Line Profile Parameters}

We extracted the intrinsic emission-line profile parameters 
(emission-line flux, FWHM, and central wavelength)
from the COS and STIS spectra using the COSFIT-freedom.pro program written 
by Kevin France for the COS Science Team. We used this extraction
procedure, as described by \citet{France2010a}, to fit either single 
and double Gaussians to the line profiles. 
Both Gaussians were fitted empirically and convolved with the point spread
function (PSF). We imposed no constraints on the central velocities of 
the Gaussians because previous solar and stellar studies show that both 
Gaussians are typically Doppler shifted from the photospheric radial velocity.
We then converted 
the measured flux $f$ (units: ergs cm$^{-2}$ s$^{-1}$) to surface flux $F$ by 
$F=f (d/R)^2$. Values for the radius ($R$), distance ($d$), and stellar 
rotation period ($P_{\rm rot}$) are given in Tables 1 and 2. 

Figure 2 shows examples of the line-fitting procedure for 
several of the emission lines of HII314, $\kappa^1$~Cet and HD~209458.
The dotted (red) lines are the intrinsic narrow and broad Gaussians, and the 
solid (blue) line is the convolution of the two Gaussians with
the COS line-spread function that best fits the observed line profile. 
Although this procedure assumes that the intrinsic line profiles are single 
or double Gaussians in shape, the fits to the observed line profiles are
generally very good even in the line wings. 
Note that the derived central wavelengths of the two Gaussians 
are not the same in these examples. Tables A1--A13 (available in the on line
journal) list the laboratory  
wavelengths \citep{Morton1991,Pagano2004}
and derived-intrinsic line parameters  
for all measured emission lines from the stars. The COS and STIS data for 
$\chi^1$~Ori are listed in separate tables. Also listed
are the heliocentric velocity shifts of the intrinsic Gaussian centroids 
compared to the laboratory wavelengths.

Since the STIS instrument was designed and calibrated 
to have an accurate velocity scale,
the STIS velocity measurements are more accurate than for COS, which has a 
large aperture with no slit. For the  STIS E140M grating data,
the scatter in measured-velocity centroids for the
photospheric lines of $\kappa^1$ Cet relative to the 
linear least-squares fit is 
about 1.5~\kms, and the velocity slope between 1300 and 1400~\AA\ is less
than 0.7 km~s$^{-1}$ and likely zero. The photospheric radial velocity of 
$\kappa^1$ Cet is listed in SIMBAD as $19.9\pm 0.9$~\kms, as 
compared to the least-squares fit to the photospheric lines near
1350~\AA\ of 17.2~\kms.
The photospheric radial velocity for $\chi^1$~Ori measured from the neutral 
lines in the STIS data near 1350~\AA\ is $-15.77\pm 0.17$~\kms 
(see Table 4) as compared to the
radial velocity listed in SIMBAD of $-13.5\pm 0.9$~\kms. We
adopt the Cl~I $\lambda$1351.657 line as our photospheric-velocity fiducial
mark because it is a bright emission line well separated from possible blends,
it is close in wavelength to the Si~IV lines, and its measured radial velocity
in the STIS $\chi^1$~Ori data (see Table 4) is $-15.51\pm 0.35$ \kms, 
close to the mean velocity of the other photospheric lines.

Since the COS instrument was designed for maximum throughput at moderate 
spectral resolution but not with a precise wavelength scale, 
there could be wavelength-dependent velocity shifts in addition to velocity
measurement errors for weak emission lines resulting from the modest 
spectral resolution. 
By comparison, STIS was designed to have an accurate
velocity scale that was calibrated in the initial STIS data pipeline 
and then further refined by the StarCAT project \citep{Ayres2010}.

To assess the wavelength dependence of the COS velocity scale in the important
1300--1400~\AA\ region, which contains emission lines of Cl~I, O~I, C~II, 
Si~IV, Fe~XII, and Fe~XXI, we plot in Figure 4 the velocity difference 
between the COS and STIS measurements of the nine brightest emission lines 
that have COS velocity-measurement errors 
(random not systematic) of less than 2.0~\kms. We have not included blended
lines, the C~II $\lambda$1334 line that has intersellar absorption, 
or the O~I $\lambda$1302, 1304, and 1306 lines that have strong 
geocoronal emission and interstellar absorption features.
Figure 4 shows a $\sim 10$~\kms\ linear
increase in the COS velocity scale between 1300 and 1400~\AA.
The flux-weighted linear fit to these data is 
vel(fit)$=-148.392+0.113988\lambda$. 
To correct for this slope in the COS velocity scale,  
we subtract 5.45~\kms\ from the COS v(Si~IV)--v(Cl~I) measurements
to force consistency between the COS and STIS velocities for $\chi^1$~Ori, 
the one star in common to both data sets. We assume that this
calibration also applies to the other COS spectra, which are mostly SNAP
observations observed with the same instrumental settings. 
We believe that this correction of the COS velocity scale in this wavelength
interval should be the same for all of the COS observations because 
the geometry of the detector should not change as the high voltage was the 
same for all observations, and thermal drifts should be largely compensated by
wavelength calibration lamp observations.

\subsection{Coronal Emission Lines}

In their search for forbidden coronal emission lines in STIS spectra of 
late-type stars and binary systems, \citet{Ayres2003} detected the Fe~XXI
$\lambda$1354.10 and/or Fe~XII $\lambda$1242.00 lines in 19 stars, but they
detected no other coronal emission lines.  The Fe~XXI and
Fe~XII lines were both detected
in the active solar-mass stars $\chi^1$ Ori and $\xi$ Boo A and 
in several dMe stars. We therefore searched the COS spectra for coronal 
emission lines. We detected the Fe~XXI line in HII314,
EK~Dra, $\pi^1$~UMa, and $\chi^1$~Ori, and the Fe~XII $\lambda$1242.00 
line in $\zeta$~Dor, $\pi^1$~UMa, $\chi^1$~Ori, and HD209458.
We also detected the Fe~XII $\lambda$1349.36 line in $\pi^1$~UMa and
$\chi^1$~Ori. As noted by \citet{Ayres2003}, there is a weak C~I line at 
1354.288~\AA\ (+41.6 km~s$^{-1}$ relative to the Fe~XXI line) that must be 
unblended to obtain the most accurate velocity and FWMH of the Fe~XXI line. 
We have made two-Gaussian fits to the blended Fe~XXI and C~I lines 
for EK~Dra, $\pi^1$~UMa, and $\chi^1$~Ori.
With increasing rotation rate, the Fe~XXI line becomes stronger relative to 
the C~I line until for HII314 the C~I line is no longer measurable.
For this star we fit the Fe~XXI line with a 
single Gaussian. Figure~3 shows the observed and fitted profiles of the Fe~XXI 
and Fe~XII lines. We also measure $3\sigma$ flux upper limits for the 
nondetected Fe~XXI and Fe~XII lines.

\section{EMISSION LINE REDSHIFTS}

\citet{Achour1995} and \citet{Peter1999} have summarized 
the observations of solar emission lines by many spacecraft
that show line-centroid redshifts of UV emission lines relative to 
photospheric lines. The disk-center redshifts of emission 
lines formed in the solar chromosphere and transition region increase
with line formation temperature up to $\log T\approx 5.3$ and then 
decrease rapidly with rising temperature. The ions with the largest redshifts
are N~V, O~IV and O~V formed at $\log T =$~5.2--5.4. \citet{Achour1995} showed 
that in solar active regions, the Si~IV and C~IV lines show redshifts that are 
about 7 km~s$^{-1}$ larger than in quiet regions, indicating that the stronger 
magnetic fields in active regions are somehow enhancing the physical processes
that produce the redshifts. There are a few measurements of stellar
emission-line redshifts \citep[e.g.,][]{Wood1997, Pagano2004, Ayres2010}, 
but until now there has been no systematic study of the redshift 
phenomenon in solar-mass stars.

In Table 5 we list the measured redshifts of the Si~IV and C~IV doublets
for the 13 solar-mass stars for which we have COS and STIS data. 
The Si~IV velocities are flux-weighted averages of the central velocities of 
single-Gaussian fits to the Si~IV $\lambda$1393 and $\lambda$1402 lines, and
the C~IV velocities are corresponding flux-weighted averages for the 
C~IV $\lambda$1548 and 1550 lines. These 
velocities are measured relative to the Cl~I $\lambda$1351 line, 
which we treat as representative of the photospheric velocity 
(see Section 2.4) 
to correct for errors in centering the star in the COS entrance aperture. 
Redshifts of the Si~IV and C~IV lines of $\alpha$~Cen~A were measured
differently. For each doublet, double-Gaussian profiles were constructed to 
best fit both lines together. The velocities listed in Table~5 for this star
are flux-weighted averages of the broad and narrow Gaussians so as to be 
comparable with the other data in the table that are single-Gaussian fits.

Velocity errors are difficult to 
quantify as they consist of both measurement errors resulting from limited 
S/N, line asymmetries, and systematic errors that result from the imperfect
instrument velocity scale and errors in the laboratory wavelengths. 
A rough estimate of the
velocity errors is the dispersion in the measured centroid velocities of the 
$\lambda$1393 and 1402 lines. The mean of this dispersion is 2.2~\kms\ for the
COS data and 1.4~\kms\ for the STIS data. The standard deviation of the 
differences between the v(COS) -- v(STIS) data and the weighted 
least-squares fit for the nine lines in Figure~4 is 1.87 \kms. This 
standard deviation provides a rough measure of the velocity errors for
the high and low S/N lines in the COS data.

\section{DISCUSSION}

\subsection{C IV and Si IV Redshifts vs. Stellar Rotational Periods}

Figure 5 shows a systematic decrease in the C~IV and Si~IV line redshifts 
measured relative to the Cl~I reference line
with increasing rotational period. This redshift-rotation correlation 
is indicated by linear least-squares fits to the STIS Si~IV and C~IV data 
separately and to the fit to the STIS and COS Si~IV data taken together. 
All three fits show systematic decreases in redshift with rotational period,
although the COS data show larger scatter about the fits. The mean errors for
each data set are 2.3 \kms\ for the COS Si~IV redshifts, 1.5 \kms\ for the
STIS Si~IV redshifts, and 1.3 \kms\ for the STIS C~IV redshifts.

Most measurements of solar UV emission lines are based on single-Gaussian fits
to emission lines observed near disk center. We include in Table~5 
the C~IV and Si~IV redshift measurements of \citet{Achour1995} obtained
with the High Resolution Telescope and Spectrometer (HRTS) instrument. These
are measurements of a quiet and an active region near disk center. We find that
the Si~IV redshift of the disk center quiet Sun (+5.4 \kms) is similar to that
of $\alpha$~Cen~A (+4.7 \kms) and the more slowly rotating stars (HD~209458 
and HD~73350), while the Si~IV redshift 
of the disk center active Sun (+11.0 \kms) is similar to that of EK~Dra
and rapidly rotating stars. C~IV redshifts of quiet and active regions near
disk center for the Sun (+6.2 and +13.0 \kms) are also similar to the
C~IV redshifts of the slowly rotating and rapidly rotating solar-mass stars.

Since the stellar data are averages over the stellar disks including limb
darkening or brightening and viewing bulk motions in the stellar atmosphere 
for all lines of sight, we need to estimate what the redshifts would be for
the Sun viewed as a star. We refer to solar disk averages as ``Sunstar'' data. 
\citet{Peter2006} presented full-disk line profiles of the C~IV $\lambda$1548
line obtained by summing raster scans of the whole solar disk 
obtained with the SUMER instrument on SOHO. The redshift
of a single-Gaussian fit to the full disk C~IV $\lambda$1548 line is +1.9~\kms,
whereas the corresponding fit to the disk-center line in the same data set 
is +4.6~\kms. He argued that
the factor of 2.4 decrease in the redshift between solar disk center and
Sunstar is due largely to viewing the mainly radial solar internal motions
tangentially at the limb. Since there are no Sunstar data for the Si~IV lines
or for active regions covering the whole Sun, we assume that this factor
of 2.4 also applies to the C~IV and Si~IV lines in the quiet and active Sun.
We list in Table~5 and plot in Figure~5 Sunstar active and quiet redshifts
estimated by dividing the disk-center redshifts of \citet{Achour1995} by this
factor of 2.4.

Figure 5 shows that the linear fit to the stellar C~IV data for the Sun's 
rotation period predicts a redshift of +4.0 \kms, which lies somewhat 
above, but within measurement errors, of the \citet{Peter2006} C~IV Sunstar 
measurement (+1.9~\kms) and consistent with the estimated redshifts for the 
C~IV lines (+2.6~\kms) and Si~IV lines (+2.3~\kms) based on the
disk-center redshifts (see Table~5) divided by 2.4. 
The Sunstar C~IV redshift velocity also lies below the 
observed +4.6 \kms\ 
redshift for $\alpha$~Cen~A and the 4.77 \kms\ redshift of the
12.3 day period star HD~73350, although the difference is likely within
measurement errors.
\citet{Peter2006} suggested that larger redshifts seen in $\alpha$~Cen~A
could result from the 25\% lower surface gravity of this star compared to the 
Sun. The estimated redshift for the active region of Sunstar corresponds to 
solar-mass stars with $P_{\rm rot}\approx 10$ days. 

Which comparison of solar to stellar redshifts is more realistic? The answer 
depends upon details of how the redshifts, which result from the heating and 
mass motions (driven by braiding of magnetic flux) in the MHD model 
of \citet{Peteretal2006}, scale with increasing
magnetic flux in the more rapidly rotating stars, and whether the the limb 
darkening/brightening behavior remains 
the same as solar for the more rapidly rotating stars that presumably are 
more covered by active regions. The answers to these questions require 
detailed models for the more rapidly rotating solar-mass stars, but we can 
now say that the magnitudes of the redshifts seen in solar-mass stars with 
rotation periods between 1.47 and 28 days are already seen on the Sun.

\subsection{Redshift Pattern vs. Stellar Rotational Period}

We plot in Figure~6 emission-line redshifts as a function of line-formation 
temperature \citep{Pagano2004}. The STIS velocities have no 
additional corrections, but the COS velocities are corrected for the 
linear slope of the COS velocity scale as described in Section 2.4. 
There is uncertainty in the laboratory wavelengths of the N~V lines 
\citep{Achour1995}. The redshifts assuming the laboratory wavelengths of 
\citet{Hallin1966} are systematically low as compared to emission
lines formed at similar temperatures, but are systematically high using the
\citet{Edlen1934} wavelengths. We have therefore adopted the mean of the two
measurements, which raises the N~V redshifts systematically by 2.74~\kms.

Figure 6 shows the redshift vs. temperature pattern relative to the 
photosphere represented by the Cl~I $\lambda$1351 line for the
two fastest-rotating stars observed with COS, four more
slowly-rotating stars observed with STIS and COS, $\alpha$~Cen~A, and, for
comparison, the solar disk-center data. 
All of the stars show a similar pattern 
of small redshifts in the chromosphere (O~I), peak redshifts at 
line-formation temperatures somewhat above $10^5$~K (e.g., C~IV and N~V) and, 
except for HII314, no significant redshifts in the Fe~XII and Fe~XXI
coronal lines. Typical measurement errors relative to the 
Cl~I line are 0.3--0.8 \kms\ for the bright lines (e.g., O~I, C~II, Si~III, 
Si~IV, and C~IV) and 1.5--3.0 \kms\ for the faint lines (e.g., O~IV and O~V)
for the stars observed with STIS and EK~Dra and $\pi^1$~UMa observed with
COS. The faintest star observed with COS, HII314, shows larger measurement 
errors of 1.3--1.7 \kms\ but 3.2 \kms\ for the faint O~V $\lambda$1371 line.
Measurement errors 
for the faint Fe~XII $\lambda$1242 line are 2.5--3.7 \kms. For the Fe~XXI
line the errors are about 2.5 \kms\ except for HII314 where the error is 
about 5 \kms. Thus, except for HII314, the redshifts of the coronal lines are 
consistent with the radial velocity of the stellar photosphere.

A remarkable result shown in Figure~6 is that all of the solar-mass stars in 
our sample, except for HII314, show very similar redshift patterns in both  
magnitude and dependence on line formation temperature
despite the wide range of $P_{\rm rot}$ between
2.61 to 28 days and the resulting very different magnetic heating rates. 
For these stars the peak redshift is between 6.0 and 10.6 \kms\ occuring 
in the temperature range $\log T = $4.8 (Si~III) to 5.25 (N~V). EK Dra and 
$\zeta$~Dor have peak redshifts slightly above those of the more 
slowly-rotating stars. The coronal line redshifts for these stars are all 
consistent with 0 \kms. 

The peak redshifts for HII314 are much larger (14.2--16.26 \kms)
than for the other stars and occur over the much larger temperature range 
$\log T = $ 4.62 (C~II) to 6.95 (Fe~XXI). This large qualitative difference 
between HII314 and the other stars is likely a consequence of the faster 
rotation of HII314. These results for HII314 suggest that for solar-mass stars 
rotating faster than about a two-day period, a qualitative break occurs 
from the usual solar-stellar 
connection in which increased magnetic heating increases the emission from the 
chromosphere and corona but not the kinematics or physical processes 
responsible for the redshift pattern. The different redshift pattern of 
HII314 suggests the appearance of different 
physical processes not seen in the slower-rotating solar-mass stars and 
thus the high activity end of the ``solar-stellar connection''.

We compare the stellar redshift results with the solar disk-center redshifts
obtained by \citet{Peter1999} using the SUMER instrument on the SOHO 
spacecraft. Since these are disk-center data rather than Sunstar data, the 
amplitudes of the redshifts are likely smaller when averaged over 
the solar disk.
If the C~IV $\lambda1548$ line is a reliable guide, the Sunstar redshifts
may be roughly a factor of 2.4 smaller than the disk-center redshifts. The
disk-center pattern of redshift vs. line-formation temperature is similar 
to that 
seen in all of the stars except HII314. The detection of solar blueshifts in
several coronal lines identifies the switch from redshift to blueshift near
$\log T \approx 5.7$ as the Ne~VIII ($\log T = 5.81$) and Mg~X 
($\log T = 6.04$) lines show small blueshifts, $-2.7\pm 1.1$ and $-4.5\pm 1.3$,
respectively \citep{Peter1999}. The measurement errors in the stellar Fe~XII 
and Fe~XXI redshifts are too large to say whether the stellar lines have 
zero or slightly negative velocities. 

\subsection{Redshifts of the Broad and Narrow Components}

\citet{Wood1997} analyzed high-resolution GHRS spectra of the Si~IV and C~IV 
lines of 12 stars including dwarfs, giants, spectroscopic binaries, 
and the Sun with spectral types F5 to M0. They found that (i) the observed 
line profiles could not be fit well by a single Doppler component, but 
could be fit better with both a narrow and a broad component, 
(ii) the fraction of the total line flux in the broad component increases 
with stellar activity, as measured by the X-ray surface flux, 
(iii) the central velocity of the narrow component is redshifted relative to
the photospheric radial velocity, and (iv) the central velocity of the
broad component can be either blueshifted or redshifted relative to the 
narrow component for different stars. They were
concerned, however, that these results and the possible explanations that 
they explored could be compromised by the range of stellar properties (mass,
spectral type, gravity, binarity, rotational  velocity) in their sample
of stars. They stated that ``ideally one should observe stars with the same 
spectral type, but different activity levels''. Our sample of stars meets 
this criterion.

We plot in Figure~7 the velocity differences between the broad and narrow 
component centroids of the two Gaussian fits to the C~IV $\lambda1548$ and
Si~IV $\lambda1393$ lines for the stars observed by STIS. We do not include 
velocity differences measured with COS, because the STIS data have higher
quality for this critical measurement as a result of higher S/N due to
their long exposure times
and more precise knowledge of the line-spread function. The mean velocity 
difference for the C~IV line observed in stars with $P_{\rm rot}$ between 
4 and 8.77 days is 5.57 \kms, with no discernable variation with
$P_{\rm rot}$. Remarkably, this mean velocity is essentially the same as
the Sunstar value of 5.1 \kms\ measured by \citet{Peter2006}, despite the 
Sun's much slower rotation. The corresponding mean value for the Si~IV
lines is 1.83 \kms\ with somewhat larger scatter. 

Inclusion of the new STIS observations of $\alpha$~Cen~A changes our 
interpretation of the velocity difference data. The new data are consistent 
with the \citet{Pagano2004} measurements but more precise: $v_b - v_n = 
-1.1\pm 0.3$ \kms\ for the Si~IV doublet and $-1.9\pm 0.3$ for the C~IV
doublet. The least squares linear fits to the velocity difference data 
including $\alpha$~Cen~A (solid lines in Figure~7) show strong increases
to shorter rotation periods. We consider these fits more realistic than
the mean values for stars in the limited rotation period range 4.0 to 8.77
days because they include the excellent $\alpha$~Cen~A data. However, 
the Sunstar C~IV velocity difference (+5.1 \kms) is inconsistent with the
$\alpha$~Cen~A data and the fits to all of the data. Our conclusion that
the velocity differences increase with decreasing rotational period
should be tested by new STIS observations
of other slowly-rotating solar-mass dwarf stars and new measurements of the
C~IV and Si~IV velocity difference for the Sun viewed as a star (Sunstar).
It is also possible, as suggested by \citet{Peter2006}, that the negative
velocity differences observed in $\alpha$~Cen~A might result
from the lower gravity of $\alpha$~Cen~A compared to the Sun and 
possible differences in its convection pattern. 

The mean fraction of the total flux in the broad component of the C~IV line
for the STIS stars is $0.52\pm 0.04$ with a range of 0.37 to 0.65. 
The Sunstar value is 0.48, consistent with the mean value.

\subsection{Physical Models for Redshifts}

The measured redshifts of UV emission lines of solar-mass stars are consistent
with the pattern shown in solar disk-center data \citep{Peter1999} with 
peak redshifts shown in lines formed near $\log T = 5.3$ and decreasing 
redshifts for lines formed at higher temperatures. The Doppler shifts for 
stellar coronal lines are consistent with either no shifts or small 
blueshifts as are seen in the solar data. This pattern is seen in all of 
the stars except for the most rapidly-rotating star HII314. On purely 
observational grounds, \citet{Peter1999} argued that the best explanation 
for the solar Doppler shift pattern involves heating below the corona that 
produces waves or flows that propagate downwards into the transition region 
and upwards into the corona. Since gas confined in magnetic loops is a 
major source of  solar UV and X-ray emission, the heating in loops likely 
occurs in one leg causing coronal gas to propagate upward along the loop,  
subsequently cool, and rain down as redshifted transition region plasma 
along the other leg of the loop. This type of explanation is now generally 
accepted for the solar data, and we propose that it is also valid for 
solar-mass stars. 

A number of authors have proposed theoretical models to explain the observed 
solar Doppler shift pattern. We cite here two recent, but very different, 
models. \citet{Zacharias2011}
developed a three-dimensional MHD model of the solar upper chromosphere and 
corona in which relatively cool chromospheric gas is pushed 
upwards and slowly heated by energy deposited at magnetic loop footpoints.
This blueshifted radiation from the upflowing gas is relatively small compared
to the redshifted radiation from the heated gas as it descends from the corona 
through the 
transition region with larger emission measure. This model, which is 
conceptually similar to the spicular model of \citet{Athay1982}, fits the
average Doppler shifts of emission lines formed in the temperature range
$\log T =$ 4.5--5.5, but does not explain blueshifts of coronal emission lines.

The three-dimensional MHD solar models computed by \citet{Hansteen2010}
differ from the \citet{Zacharias2011} model in that the heating occurs 
primarily in the transition region by rapid, episodic events. The 
heating-induced overpressure produces downflows of the underlying plasma
and upflows in the overlying coronal plasma. The effect is to produce
redshifts of transition region lines like C~IV and no or small blueshifts
of coronal lines like Fe~XII and Fe~XV, consistent with solar-average line
shifts and the stellar redshift data. 
An interesting result appears in their model for weak photospheric
magnetic fields and minimal heating. They find no redshifts in this model and
conclude that ``transition region redshifts are a direct consequence to 
vigorous heating to coronal temperatures at low heights, with redshifts 
disappearing as the amplitude of the heating decreases.'' Since the UV 
emission-line fluxes (see Fig. 1), and thus heating, increase with rapid 
rotation,
the trend of increasing Si~IV and C~IV redshifts with rapid rotation is 
in qualitative agreement with the \citet{Hansteen2010} models. We note that
our estimate of the C~IV redshift for a solar-mass star completely covered 
with solar network elements with their enhanced heating is similar to the 
observed redshift of EK Dra ($P_{\rm rot}=2.6050$ days), a solar-mass star with
nearly 100 times the flux (and thus heating) in the Si~IV lines. While
additional calculations are needed to tie down the detailed connection
between heating, transition-region line emission, and redshifts, the
\citet{Hansteen2010} calculations indicate that the heating mechanism in their 
models could explain both phenomena.

\citet{Wood1997} suggested that the broad component of emission lines formed in
the chromosphere and transition region is produced by large motions associated 
with explosive events that may result from nanoflare heating. \citet{Peter2001}
proposed a very different model in which the broad components are produced by 
upward-propagating magneto-acoustic waves in coronal funnels.
In the three-dimensional
MHD models discussed by \citet{Hansteen2010}, transient heating in the 
transition region occurs largely in 
spatially restricted regions where the added pressure produced by rapid 
heating generates high-speed mass motions. The sum of such events over the 
stellar surface could explain the observed broad components, in which
case the fractional flux in the broad components is a measure of the 
transient-heating process.

\section{CONCLUSIONS}

In this systematic study of ultraviolet emission lines formed in the outer 
atmospheres of solar-mass dwarf stars, we investigate how the parameters of 
these emission lines
depend upon a stellar rotation period between 1.47 to 28.0 days
for stars that otherwize have similar parameters 
(e.g., mass, $T_{\rm eff}$, $g$, chemical composition). 
This study is based on high-resolution ultraviolet spectra of six stars 
observed with the COS instrument on HST, eight stars observed with the STIS
instrument on HST, and the full-disk solar C~IV $\lambda$1548 line profile. 
In this second paper of a series, we find the following results concerning 
emission-line redshifts:

(1) We find that increasing redshifts are correlated with more rapid rotation
(decreasing $P_{\rm rot}$) for the Si~IV and C~IV lines. 
The quiet Sun disk center Si~IV and C~IV redshifts lie slightly above, but are 
within measurements errors of,  
this trend, and the active Sun disk center redshifts are similar to
those seen in EK~Dra and other rapidly rotating stars. The observed 
solar full disk (Sunstar) redshift for the C~IV $\lambda$1548 line lies
slightly below, but also within measurement errors of, least-squares fits to 
the other stars. New theoretical studies of how redshifts
depend on the magnitude and direction of gas flows produced by magnetic 
heating are needed to determine how best to estimate solar full disk redshifts 
from disk center observations.

(2) The redshift pattern with line-formation temperature for EK~Dra 
($P_{\rm rot}=2.61$ days) is similar to that of the  
more slowly rotating stars (including the Sun) with maximum redshift
near $T=10^5$~K and no redshift of the coronal Fe~XII and Fe~XXI lines.
This similarity indicates that the coronal plasma observed in the Fe~XII and 
Fe~XXI lines is not participating in a stellar wind but is instead constrained 
by magnetic fields to corotate with the stellar photosphere.
Unlike the more slowly rotating stars, the fastest-rotating solar-mass star 
in our study, HII314,
shows significantly enhanced redshifts at all temperatures above $\log T=4.6$, 
including the corona. This difference in the redshift pattern 
between HII314 and EK Dra suggests that a 
qualitative change in the magnetic-heating process occurs near 
$P_{\rm rot}=2$ days. We propose that HII314 is an example of a solar-mass star
that is rotating too fast with a magnetic heating rate that is too large
for the physical processes that control the redshift pattern to operate 
in the same way as for the more slowly rotating stars. HII314
appears to lie above the high activity end of the set of 
solar-like phenomena that is typically called the 
``solar-stellar connection''. 

(3) The stronger emission lines in these stars are usually better fit by
two Gaussians (one broad and the other narrow) rather than by a single 
Gaussian. We find that the mean velocity difference between the centroids 
of the broad and narrow components of 
the C~IV $\lambda1548$ lines observed in stars with $P_{\rm rot}$ between 
4 and 8.77 days is 5.57 \kms, with no discernable variation with
$P_{\rm rot}$. Remarkably, this mean velocity difference 
is essentially the same as
the Sunstar value of 5.1 \kms\ measured by \citet{Peter2006}, despite the 
Sun's much slower rotation rate. 
The corresponding mean velocity difference 
for the Si~IV lines is 1.83 \kms\ with somewhat larger scatter.

The inclusion of new STIS spectra of
$\alpha$~Cen~A identifies a trend of increasing velocity differences with 
more rapid rotation. This result should be tested by new STIS observations
of slowly rotating solar-mass stars and new measurements of the velocity 
differences for the Sun viewed as a star (Sunstar).
The mean fraction of the total flux in the broad component of the C~IV line
is $0.52\pm 0.04$ with a range of 0.37 to 0.65. The Sunstar value is
0.48, consistent with the mean value. The broad components could be the sum of
many transient-heating events in which the added pressure produced by 
rapid heating generates high-speed mass motions \citep{Hansteen2010}. 
If this model is correct, then
the fractional flux in the broad components is a measure of the transient
heating process. We will return to this question in a later paper in this 
series in which we discuss the observed stellar line widths using the data in
Tables A1-A13. 

(4) Redshifts result from flows produced by magnetic heating. The observed 
pattern of small redshifts in lines formed in the chromosphere, maximum 
redshifts for lines formed in the temperature range $\log T = $4.8--5.25, and 
no redshifts or small blueshifts of coronal lines is consistent with the 
predictions of the three-dimensional MHD solar model of \citet{Hansteen2010}. 
We suggest that
the \citet{Hansteen2010} model be run at very large heating rates to 
determine whether the model can reproduce the different redshift pattern
seen in HII314. This would be a good test of the applicability of the model 
for very high magnetic heating rates.

\acknowledgements

This work is supported by NASA through grants NNX08AC146, NAS5-98043, 
and HST-GO-11687.01-A to the University of Colorado at Boulder. 
We thank Tom Woods and Martin Snow
for providing the SORCE data and Steven Osterman for information on the 
calibration of COS.

{\it Facilities:} \facility{HST (COS)}, \facility{HST (STIS)}, 
\facility{SUMER/SOHO}, \facility{SORCE/SOLSTICE II},
\facility{SIMBAD}

\appendix
\section{TABLES OF EMISSION LINE WAVELENGTHS, VELOCITIES, AND FLUXES}

Tables A1--A13 (available in the online journal) 
list the emission-line parameters for the high-resolution stars 
observed by COS and STIS that are the basis for this study. The data for
$\alpha$~Cen~A are published by \citet{Pagano2004}, and the sources for 
the solar data are
given in Section 2.4. The second column in each table is the laboratory 
wavelength given by \citet{Morton1991} when available or by \citet{Pagano2004}.
The third and fourth columns give the measured wavelengths (and errors) and 
the corresponding radial velocities (and errors). The next two columns give 
the measured fluxes and FWHM of the lines with their
measurement errors. For most of the emission lines, a single Gaussian 
provides a good fit to the data. For the brightest emission lines, we found 
that two Gaussians (a narrow and a broad Gaussian) provide a better fit to
the data. For these bright lines, we list the parameters for both the narrow 
and broad Gaussian components described by the subscripts $n$ and $b$. The 
asterix superscript after the ion indicates that we computed a two-Gaussian fit
to the emission line. For the same lines, we also provide the parameters
for a single-Gaussian fit. These parameters are useful for comparison with 
fainter lines that can only be fit by a single Gaussian. Those lines marked
$bl$ are close blends of lines from the same ion for which we use the 
mean laboratory wavelength. Widely separated blends or blends involving
more than one atom or ion are not included.

We do not include very weak lines for which the flux errors exceed about 30\%
as the radial velocities of these lines are unreliable. We include flux
upper limits for the coronal Fe~XII $\lambda$1242.00 and $\lambda$1349.36 
and Fe~XXI $\lambda$1354.080 lines when no emission feature
is present. We have assumed that the FWHM = 45~\kms\ for the nondetected
Fe~XII lines and 110~\kms\ for the nondetected Fe~XXI lines.
For two stars (HII314 and HD 209458) the COS observations with the G130M and 
G160M gratings were taken at different times with the star likely placed 
in different portions of the aperture. For this reason the velocity
scales for the two gratings are different. A solid horizontal line separates 
the data obtained with the different gratings. Table A5 lists the data for
$\chi^1$~Ori obtained with COS and Table A6 lists the data obtained with STIS.

\begin{figure}
\includegraphics[angle=180, scale=0.7]
{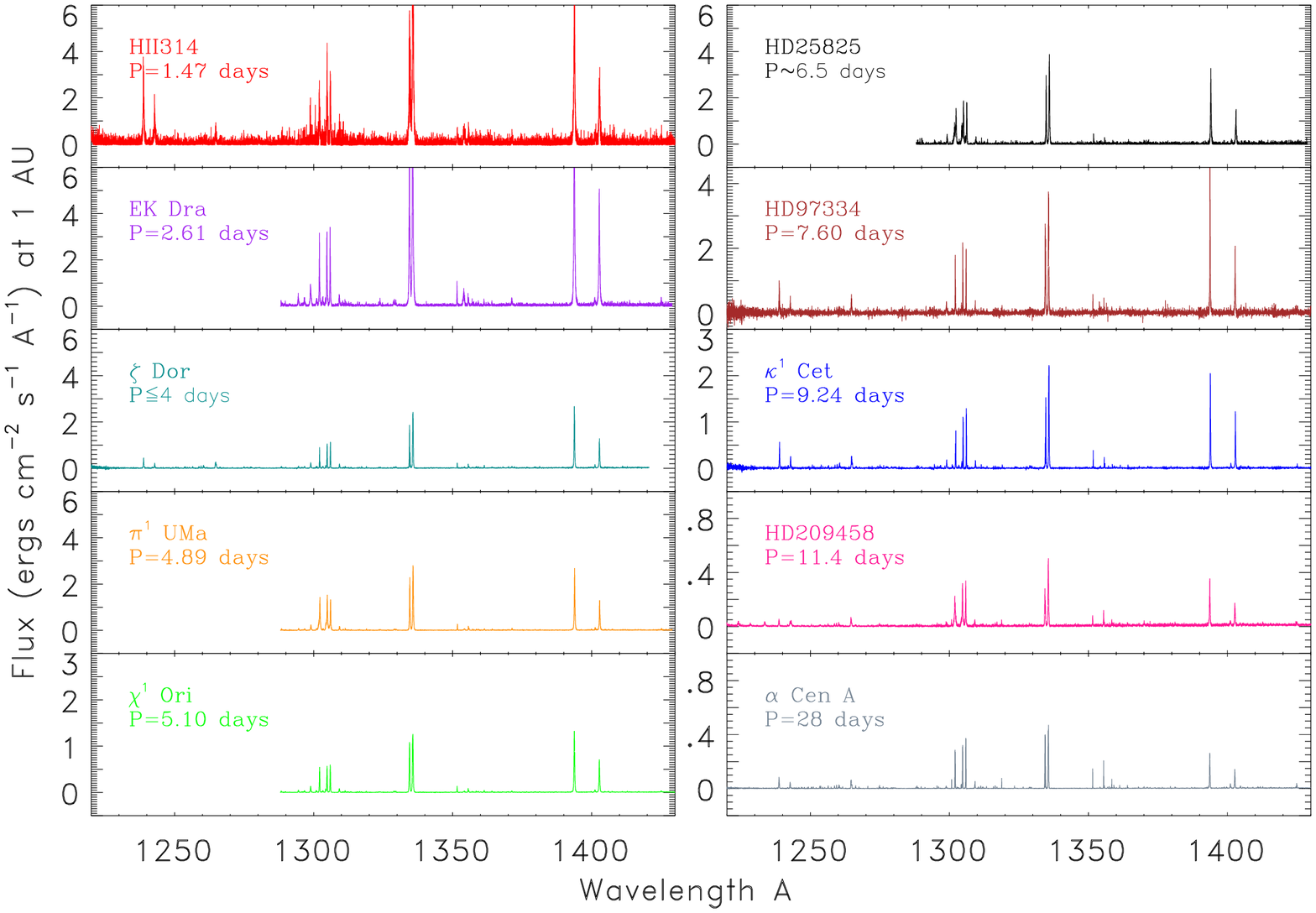}
\caption{Comparison spectra of ten of the stars discussed in this paper.
The stars are listed in order of decreasing rotation period (P). The stars 
observed as a part of the SNAP program have a more limited spectral range.
Note that the spectra of the slower rotating and thus older stars have 
different flux scales than the faster rotating stars.
(A color version of this figure is available in the online journal.)} 
\end{figure} 

\begin{figure}
\includegraphics[scale=0.7]
{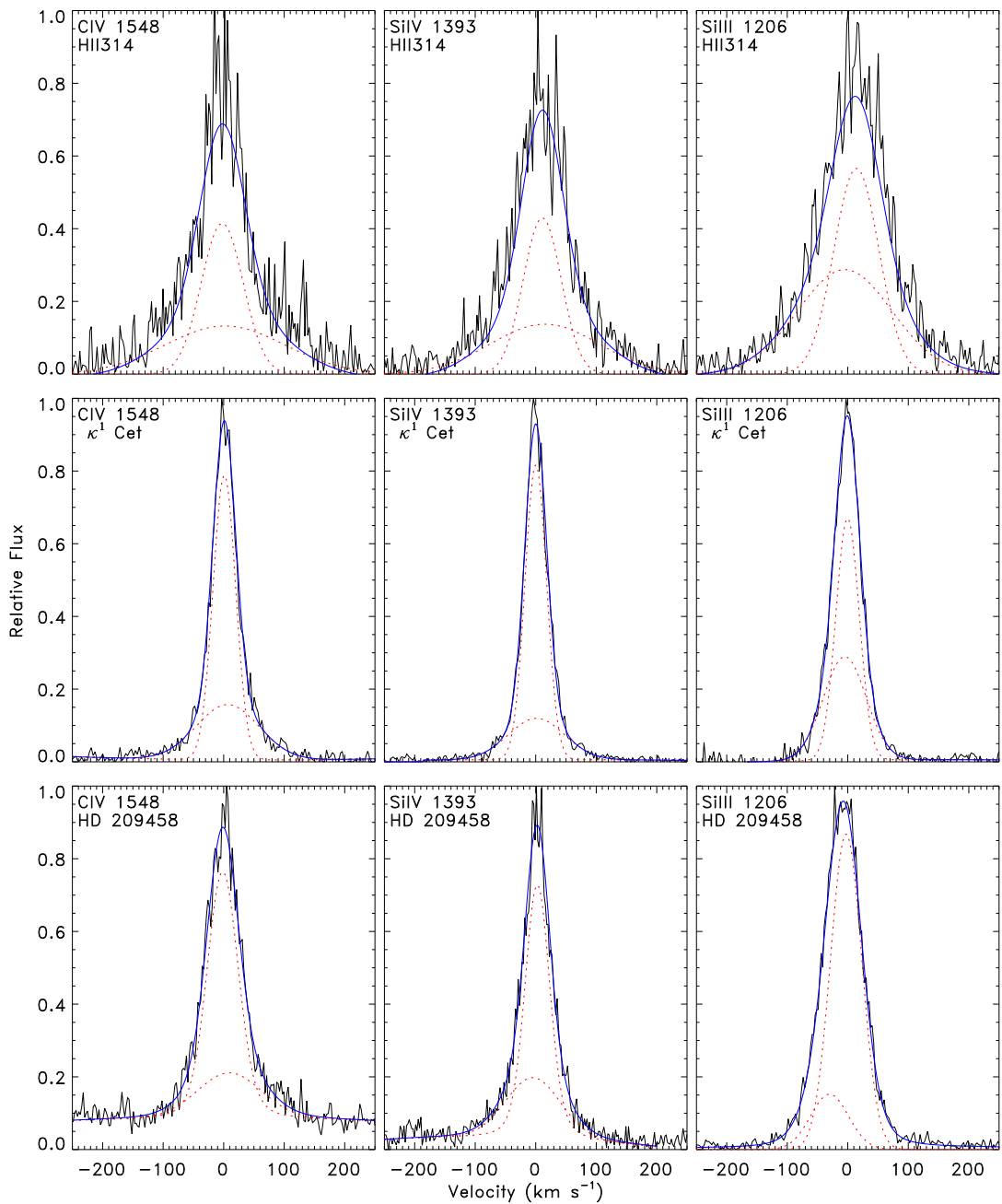}
\caption{Two component fits to the C~IV $\lambda$1548, Si~IV $\lambda$1393,
and Si~III $\lambda$1206 lines of three stars HII314, $\kappa^1$~Cet, and 
HD~209458. The narrow and broad component fits are shown as
dotted (red) lines, and the solid (blue) lines are convolutions of the two 
Gaussians with the COS line-spread function that best fits the observed line 
profile. The broad components contain a large portion of 
the line fluxes for the rapid rotating star HII314 ($P_{\rm rot}=1.47$ days) 
but much smaller portions of the line fluxes for the more slowly rotating 
stars $\kappa^1$~Cet ($P_{\rm rot}=8.77$ days) and HD~209458 
($P_{\rm rot}=11.4$ days).
(A color version of this figure is available in the online journal.)}
\end{figure}

\vspace{-2.0in}
\begin{figure}
\includegraphics[scale=0.70, angle=90]
{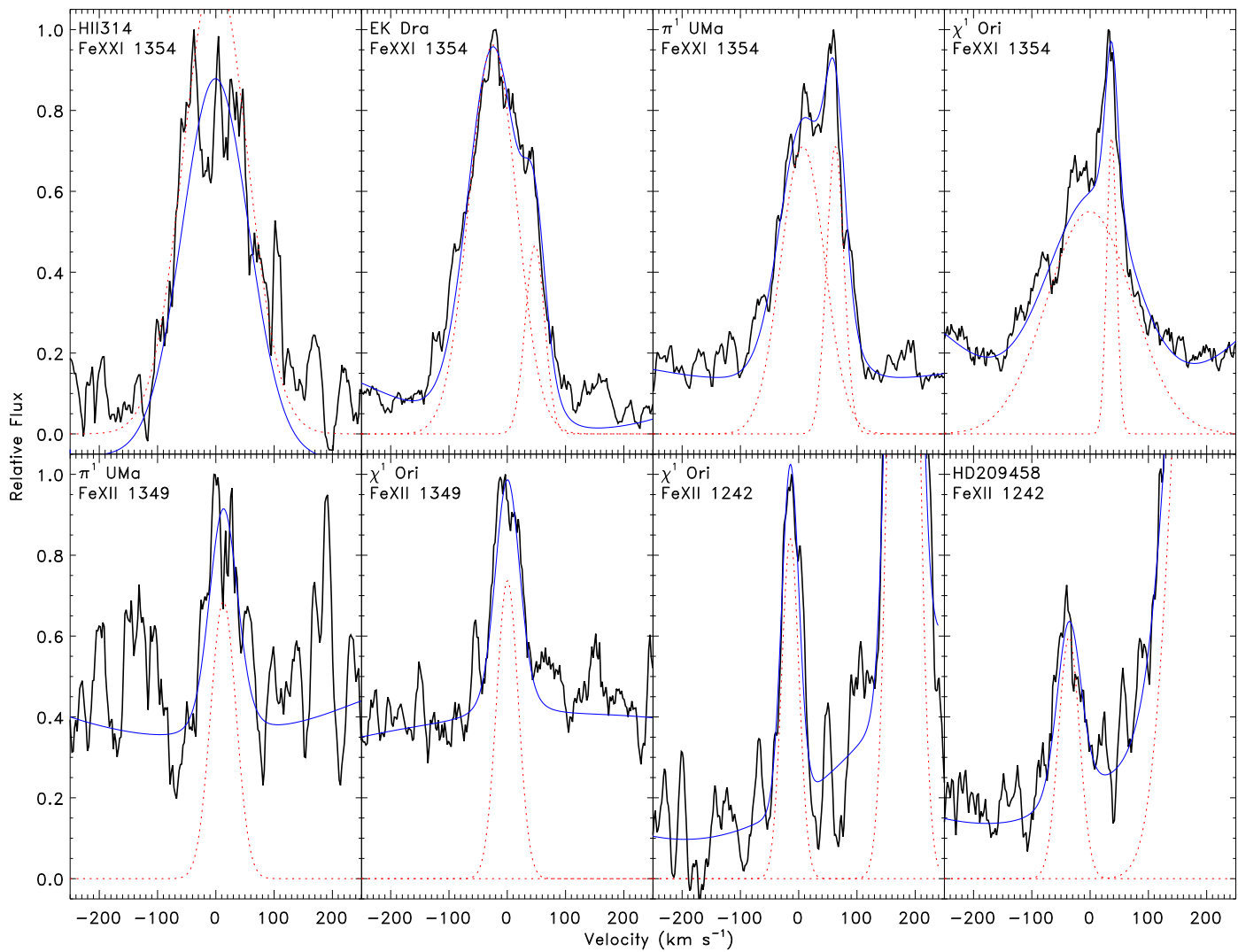}
\caption{Comparison of the Fe~XXI $\lambda$1354.080 and Fe~XII 
$\lambda$1242.000 and $\lambda$1349.36 lines observed in five stars. 
The C~I $\lambda$1354.288 line is identified and fit for
three of the stars showing the Fe~XXI line. With increasing rotation rate, the 
coronal Fe~XXI line becomes brighter relative to the photospheric C~I line.
(A color version of this figure is available in the online journal.)}
\end{figure}
\pagebreak

\begin{figure}
\includegraphics{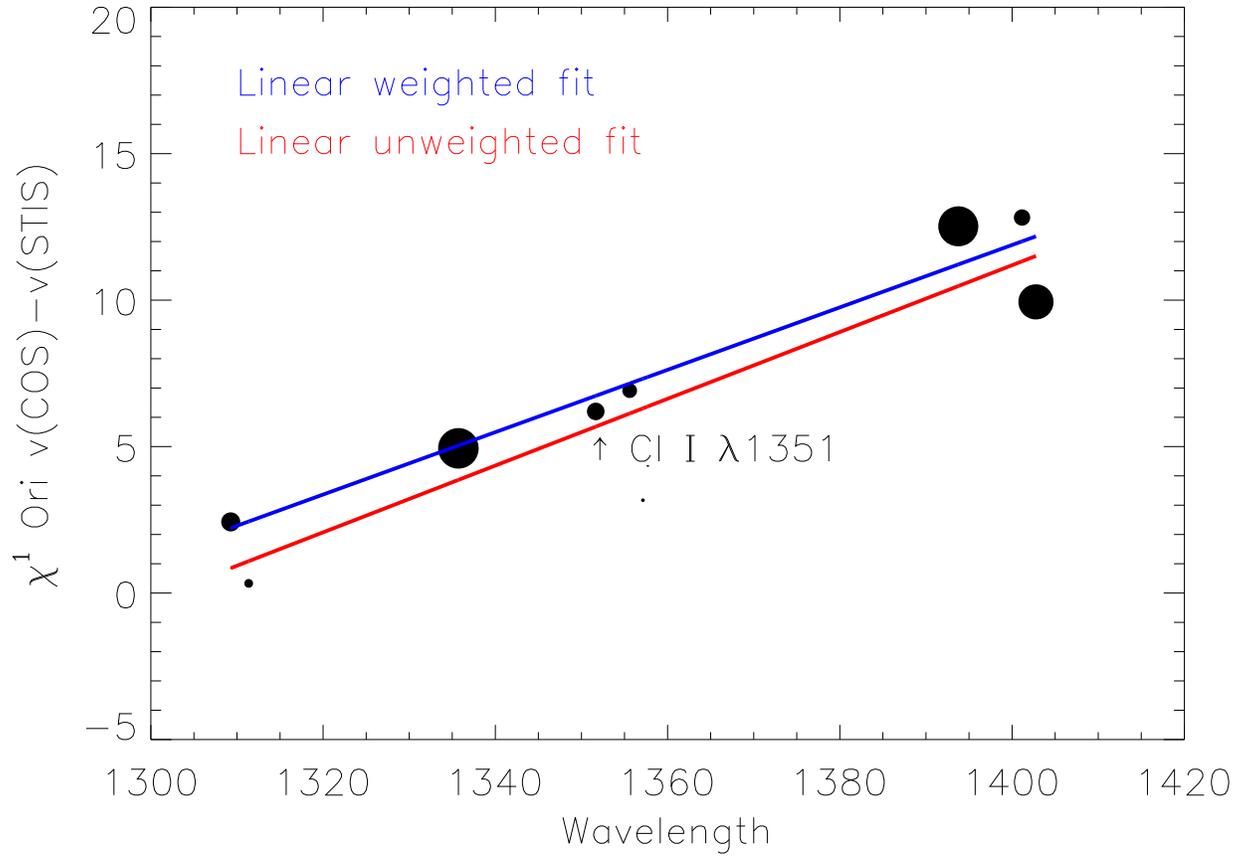}
\caption{Radial velocity differences between the $\chi^1$~Ori COS and STIS 
measurements of the 9 emission lines in common with COS velocity-measurement
errors less than 2.0 \kms. The symbol sizes are 
proportional to the log of the COS line fluxes. Also plotted are flux-weighted
and unweighted linear least-squares fits to the data.
(A color version of this figure is available in the online journal.)}
\end{figure}

\begin{figure}
\includegraphics[angle=0, scale=1.0]
{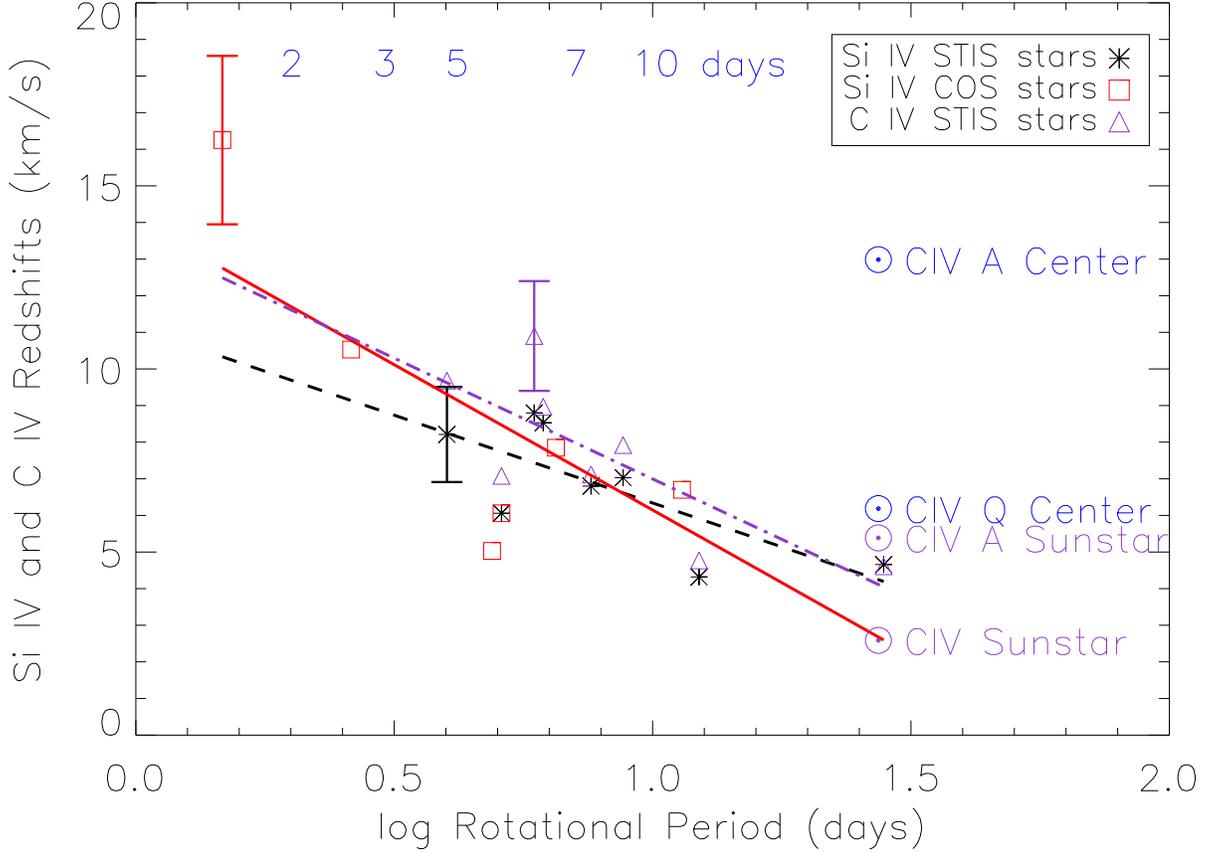}
\caption{Redshifts of the Si~IV and C~IV lines plotted vs. stellar 
rotational period
for solar-mass stars observed with the COS and STIS instruments on HST. The
plotted Si~IV velocities are flux-weighted averages of the central velocities 
of single-Gaussian fits to the $\lambda$1393 and 1402 lines, except for the 
$\alpha$~Cen~A velocities as described in Section 3. The Si~IV 
velocities are measured relative to the photospheric Cl~I $\lambda$1351
line, and the COS Si~IV velocities are shifted by --5.45~\kms\ 
to be consistent with
the STIS velocity scale for $\chi^1$~Ori (see Section 3). The C IV lines,
all measured from STIS spectra, are flux-weighted averages of the $\lambda$1548
and 1550 lines minus the Cl~I line velocity. The error bars are the mean
errors for each data set: 2.3 \kms\ for the COS Si~IV redshifts 
(larger for HII314), 1.5 \kms\
for the mean STIS Si~IV redshifts, and 1.3 \kms\ for the STIS C~IV redshifts.
The dashed (black) line is the linear least squares fit to the STIS Si~IV
redshifts, the dash-dot (violet) line is the corresponding fit to the STIS C~IV
redshifts, and the solid (red) line is the fit to all of the Si~IV redshifts. 
We include C~IV redshifts for the solar disk-center quiet (CIV Q Center) 
and active (CIV A Center) regions, for the integrated disk (CIV Sunstar), 
and for the Sun covered with active regions (CIV A Sunstar, see Section 4).
(A color version of this figure is available in the online journal.)}

\end{figure}

\begin{figure}
\includegraphics{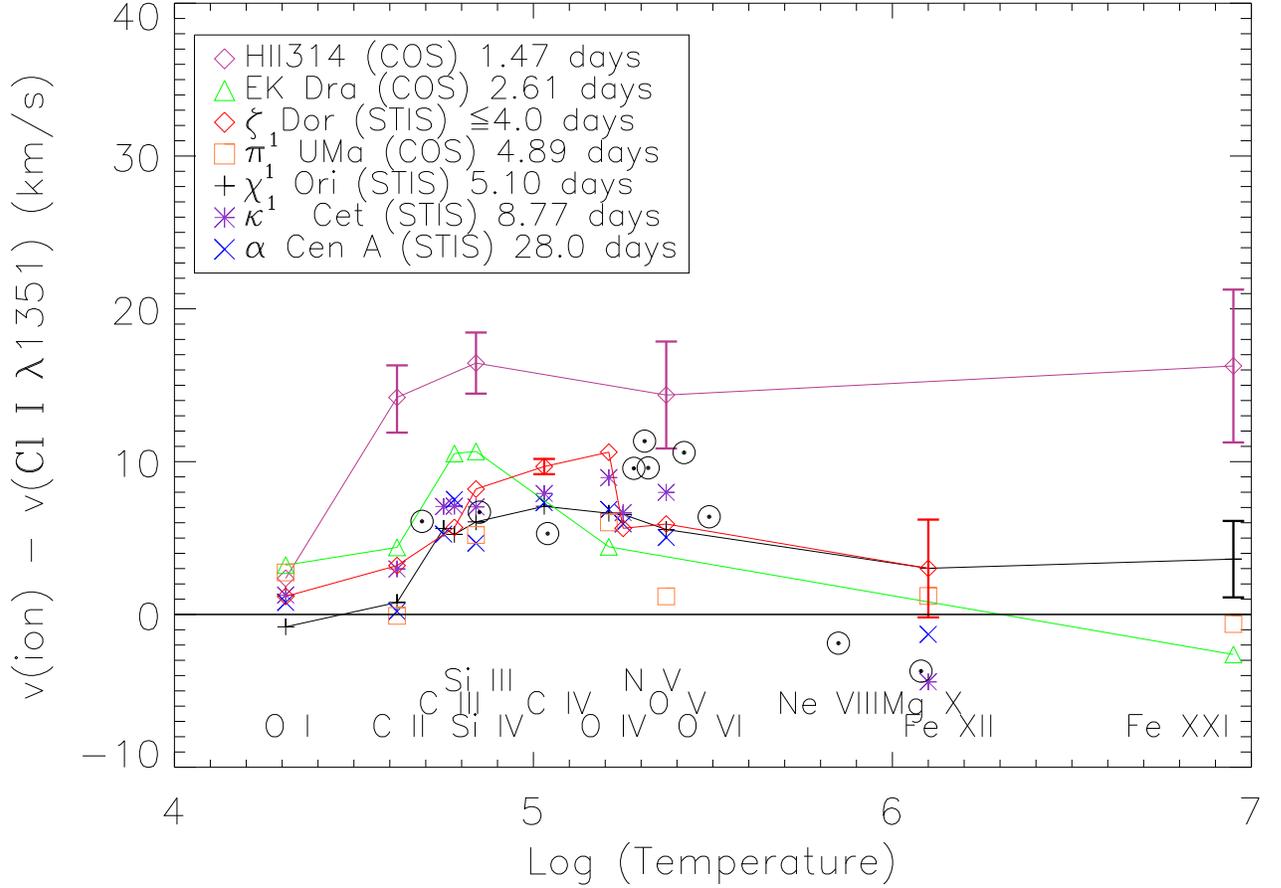}
\caption{Emission-line redshifts relative to the Cl~I $\lambda$1351 line 
representing the stellar photosphere radial velocity. 
The sources for the data are listed in 
the legend, except that the Fe~XXI velocity for $\chi^1$~Ori was 
obtained with COS. Typical measurement errors relative to the 
Cl~I line are 0.3--0.8 \kms\ for the bright lines (e.g., O~I, C~II, Si~III, 
Si~IV, and C~IV) and 1.5--3.0 \kms\ for the faint lines (e.g., O~IV and O~V)
for the stars observed with STIS and EK~Dra and $\pi^1$~UMa observed with
COS. The faintest star observed with COS, HII314, shows larger measurement 
errors of 1.3--1.7 \kms\ but 3.2 \kms\ for the faint O~V $\lambda$1371 line.
Typical measurement errors for the Fe~XII $\lambda$1242 line observed by STIS 
are 2.5--3.7 \kms\ and for the Fe~XXI $\lambda$1354 line are 2.5--5.0 \kms\
as shown in the figure.
EK Dra ($P_{\rm rot}=2.61$ days) shows a similar redshift pattern
to the slower rotators, but HII314 ($P_{\rm rot}=1.47$ days) shows an 
enhanced redshift pattern at all temperatures above $\log T = 4.6$. Redshifts 
of the coronal lines (Fe~XII and Fe~XXI) are consistent with 0 \kms\ 
within the measurement errors except for HII314. The Sun symbols show solar 
disk-center redshift measurements of \citet{Peter1999} using the SUMER 
instrument on the SOHO spacecraft.
(A color version of this figure is available in the online journal.)}
\end{figure}

\begin{figure}
\includegraphics{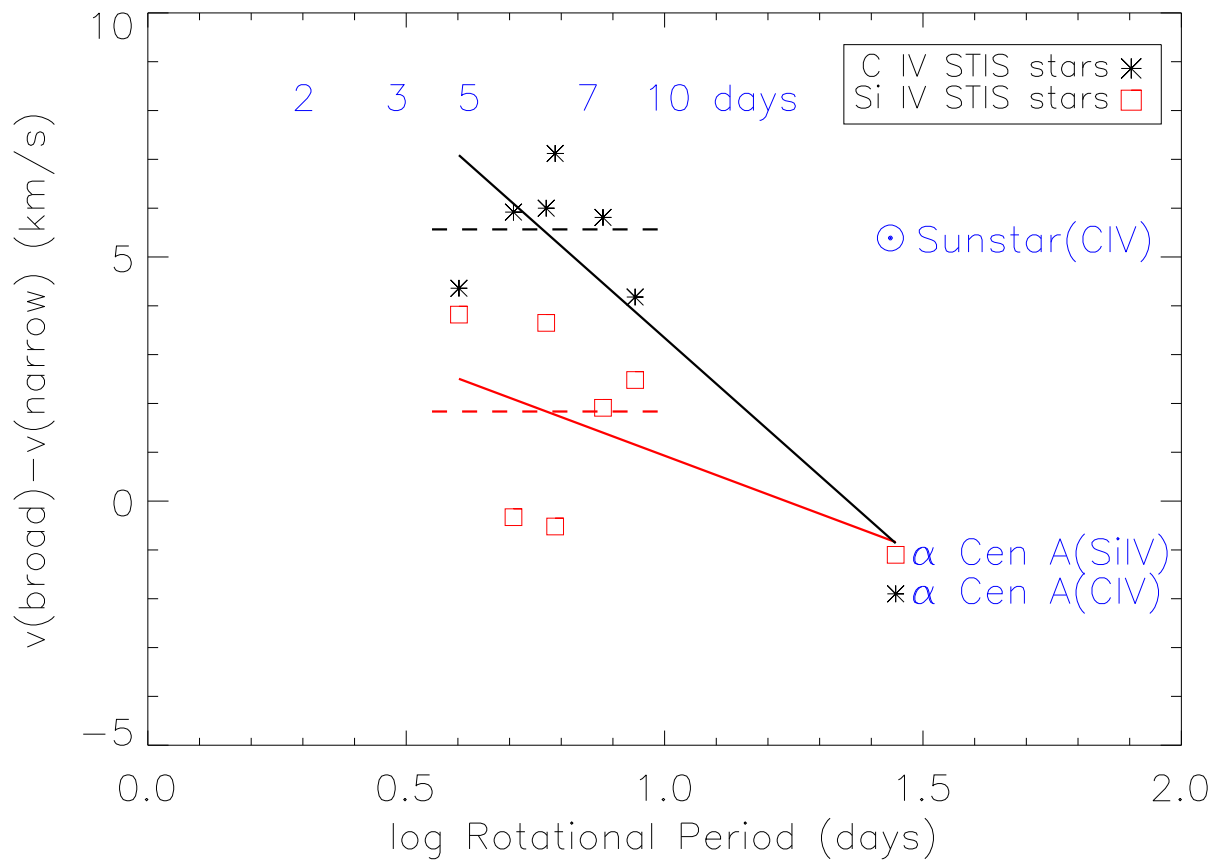}
\caption{Velocity differences between the broad and narrow component 
centroids of the two Gaussian fits to the C~IV $\lambda1548$ and 
Si~IV $\lambda1393$ lines for the stars observed with STIS. Also shown is the
velocity difference for the C~IV line of the Sun viewed as a star (Sunstar).
The mean values of the velocity differences (dashed lines) 
for all stars except $\alpha$~Cen~A
are 5.57 \kms\ for C~IV and 1.82 \kms\ for Si~IV. The Sunstar C~IV velocity
difference of 5.1 \kms\ \citep{Peter2006} closely agrees with the stellar
mean value. The solid lines are linear least-squares fits to all of the
C~IV and S~IV stellar data, including $\alpha$~Cen~A, which we believe are 
more realistic results but they should be tested by new data.
(A color version of this figure is available in the online journal.)}

\end{figure}

\begin{deluxetable}{cccccccc}
\tablewidth{0pt}
\tabletypesize{\scriptsize}
\rotate
\tablenum{1}
\tablecaption{Properties of the Solar Mass Stars Obtained with 
High Spectral Resolution}
\tablehead{Property/Star & HII314 & EK Dra & $\zeta$ Dor &
$\pi^1$ UMa & $\chi^1$ Ori & HD 165185 & HD 25825} 

\startdata
Age/$10^9$(yr) & 0.1 & 0.03--0.05 & & 0.3 & 0.3 & 0.3 & 0.6\\
Cluster & Pleiades & Pleiades MG & & UMa MG & UMa MG & UMa MG & Hyades\\
$P_{rot}$(days) & 1.47 & 2.6050 & $<4.0$ & 4.89 & 5.10 & 5.90 & $\approx6.5$\\
$vsini$(km/s) & $38.7\pm1$ & $17.3\pm 0.4$ & 14.8 & $9.7\pm 0.4$ & 
$9.0\pm 0.2$ & $7.6\pm 1.2$  & 6.8\\
Sp Type & G1 V & G1.5 V & F9V & G1.5 V & G1 V & G1V & G0 V\\
$R/R_{\odot}$ & 1.0 & 0.95 & 0.96 & 0.95 & 0.98 & 0.94 & 1.0\\
Other name & V1038 Tau & HD 129333 & HD 33262 & HD 72905 & HD 39587 & 
HIP 88694 & VB~10\\
d(pc) & 134 & 34 & 11.7 & 14.3 & 8.7 & 17.4 & 46.34\\
log $L_x$ & 30.3 & 29.9 & & 29.1 & 29.0 & 29.17 & 28.9\\
$F_x/10^6$ & 16 & 10 & & 2.0 & 2.0 & & 1.2\\ \hline
HST Spectrograph & COS G130M & COS G130M & STIS E140M & COS G130M & 
COS G130M &  STIS E140M & COS G130M\\
HST Program & 11532 & 11687 & 8280 & 11687 & 11687 & 9113 & 11687\\ 
Observation Date & 12/16/2009 & 4/22/2010 & 5/01/1999 & 2/28/2010 & 
3/19/2010 & 9/18/2002 & 2/20/2010\\
Exposure Time(sec) & 3209 & 1160 & 6000 & 1300 & 1300 & 3841 & 1160\\
Other observations & FOS,GHRS,IUE & GHRS,IUE,FUSE & STIS E230H & STIS E230H & 
STIS E140M,E230H & IUE & FOS\\
\enddata
\end{deluxetable}

\begin{deluxetable}{cccccccc}
\tablewidth{0pt}
\tabletypesize{\scriptsize}
\rotate
\tablenum{2}
\tablecaption{Properties of the Solar Mass Stars Obtained with High Spectral 
Resolution}
\tablehead{Property/Star\tablenotemark{a} & HD 97334 & $\kappa^1$ Cet & 
HD 73350 & HD 59967 & HD209458 & Sun & $\alpha$ Cen A}

\startdata
Age/$10^9$(yr) & $0.45\pm 0.20$ & 0.75 & $0.51\pm 0.14$ & $0.35\pm 0.07$ 
& $4\pm 2$ & 4.6 & 3.9--4.9\\
Cluster & & & & & & & \\
$P_{rot}$(days) & 7.60 & 9.24 & $12.3\pm 0.1$ & 6.14 & 11.4 & 25.4 & 28.0\\
$vsini$(km/s) & 5.4 & 4.5 & $4.0\pm 0.5$ & 3.8 & 4.5 & 1.7 & 3.2\\
Sp Type & G0 V & G5 V & G0  & G3 V & G0 V & G2 V & G2 V\\ 
$R/R_{\odot}$ & 1.01 & 0.99 & 1.07 & 0.89 & 1.125 & 1.00 & 1.224\\
Other name & MN UMa & HD 20630 & HIP 42333 & HIP 36515 & V376 Peg & -- & 
HD 128620\\
d(pc) & 23.8 & 9.2 & 23.6 & 21.8 & 47 & -- & 1.338\\
log $L_x$ & 29.1 & 28.8 & &  &  & 26.3--27.3 & 26.5--27.0\\
$F_x/10^6$ & 1.8 & 1.0 &  &  &  &  0.0002--0.002 & 0.003--0.01\\
\hline
HST Spectrograph & STIS E140M & STIS E140M & STIS E140M & STIS E140M & 
COS G130M & SORCE & STIS E140H\\
HST Program & 9113 & 8280 & 9113 & 9113 & 11534 & -- & 7263\\
Observation Date & 6/21/2003 & 9/19/2000 & 4/26/2003 & 9/10/2003 & 
9+10/2009 & 3/2008 & 2/12/1999\\
Exposure Time(sec) & 3841 & 7805 & 3841 & 3841 & 20694 & 180 & 14085\\
Other observations & IUE & STIS E230H & -- & -- & STIS,GHRS & SOHO/SUMER & 
GHRS,FUSE\\
\enddata
\tablenotetext{a}{Age references: Plavchan et al. (2009), 
Ribas et al. (2010).\\
$P_{\rm rot}$ and $vsini$ references: Pizzolato et al. (2003), 
Gaidos et al. (2000), Petit et al. (2008).}

\end{deluxetable}

\begin{deluxetable}{lccccccc}
\tablewidth{0pt}
\tablenum{3}
\tablecaption{Observing Log of Eight Stars}
\tablehead{Star\tablenotemark{a} & Instrument & Dataset & Grating & 
Central $\lambda$ & Yr-mn-day & Start Time &T$_{exp}$(s)}
\startdata
HII314 & COS & LB6401010  & G130M & 1309 & 2009-12-12 & 16:09:28 & ~814.1\\
HII314 & COS & LB6401020  & G130M & 1318 & 2009-12-12 & 16:09:45 & ~990.1\\
HII314 & COS & LB6401040  & G130M & 1300 & 2009-12-12 & 16:11:14 & 1405.2\\
HII314 & COS & LB6402010  & G160M & 1600 & 2010-10-11 & 11:49:04 & 1402.0\\
HII314 & COS & LB6402020  & G160M & 1577 & 2010-10-11 & 12:15:32 & ~326.9\\
EK Dra & COS & LB3E34010  & G130M & 1291 & 2010-04-22 & 09:18:42 & 1160.0\\
$\pi^1$ UMa&COS&LB3E26010 & G130M & 1291 & 2010-02-28 & 22:31:16 & 1300.4\\
$\chi^1$ Ori&COS&LB3E06010& G130M & 1291 & 2010-03-19 & 06:47:26 & 1300.4\\
$\chi^1$ Ori&STIS&O5BN02010--30&E140M&1425&2000-10-03 & 01:11:14 & 6770.0\\
HD25825& COS & LB3E41010  & G130M & 1291 & 2010-02-20 & 23:15:33 & 1160.0\\
$\kappa^1$ Cet&STIS&O5BN03050--60&E140M&1425&2000-09-19&08:58:48 & 7805.0\\
HD165185&STIS& O6CO07040  & E140M & 1425 & 2001-09-18 & 00:05:52 & 2977.0\\
HD165185&STIS& O6CO07050  & E140M & 1425 & 2001-09-17 & 23:11:23 & ~864.0\\
HD73350 &STIS& O6CO06040  & E140M & 1425 & 2002-04-26 & 03:32:25 & 2899.0\\
HD73350 &STIS& O6CO06050  & E140M & 1425 & 2002-04-26 & 02:30:59 & ~864.0\\ 
 
\enddata
\tablenotetext{a}{For information on the observations of $\alpha$~Cen~A see
\citet{Pagano2004}. For information on the observations of HD~209458 see
\citet{Linsky2010} and \citet{France2010a}.}

\end{deluxetable}

\begin{deluxetable}{lccc}
\tablewidth{0pt}
\tabletypesize{\small}
\tablenum{4}
\tablecaption{Radial Velocities for $\chi^1$ Ori 
(STIS data)}
\tablehead{Atom & Number & $<v>\pm 1\sigma$\tablenotemark{a} & 
Total flux\\
or Ion & of Lines & (\kms) & ($10^{-15}$)}

\startdata
Cl I  & 1 & $-15.51\pm 0.35$ & 10.49\\
S I   & 9 & $-15.17\pm 0.79$ & 23.18\\
Si I  & 9 & $-16.56\pm 0.76$ & 35.07\\
C I   & 21& $-15.96\pm 0.42$ & 60.50\\
N I   & 2 & $-15.04\pm 0.76$ &  5.25\\
O I   & 3 & $-16.32\pm 0.06$ &156.07\\
N II  & 3 & $-15.14\pm 3.69$ &  4.85\\
Fe II &33 & $-15.37\pm 0.47$ &218.97\\
Si II &17 & $-15.50\pm 0.55$ &169.94\\ \tableline
      &   &                  &      \\
Sum S I to Si II & 97 & $-15.77\pm 0.17$ & 673.83\\
      &   &                  &      \\ \tableline
C II  & 2 & $-14.91\pm 0.23$ &359.16\\
He II & 1 & $-11.10\pm 0.23$ &258.45\\
S III & 1 & $-11.48\pm 1.79$ & 46.70\\
Si III& 7 & $-13.69\pm 0.53$ &378.67\\
C III & 3 & $-9.87\pm 1.21$  & 72.66\\
N IV  & 1 & $-12.66\pm 2.57$ & 33.80\\
O IV] & 3 & $-10.00\pm 0.17$ &106.76\\
Si IV & 2 & $-9.45\pm 0.34$  &302.14\\
C IV  & 2 & $-7.53\pm 1.95$  &390.63\\
N V   & 2 & $-11.71\pm 0.20$ & 46.16\\
O V   & 2 & $-9.95\pm 0.29$  & 30.25\\
Fe XII& 2 & $-8.86\pm 5.29$  &  3.34\\

\enddata
\tablenotetext{a}{$\sigma$ is the standard error of the weighted-mean radial
velocity when more than one line is available or the error in the centriod
of the line profile when there is only one line.}

\end{deluxetable}

\begin{deluxetable}{lccccccccc}
\tablewidth{0pt}
\rotate
\tablenum{5}
\tablecaption{Emission Line Redshifts Using the Cl I 1351~\AA\ Line as a 
Fiducial for the Photospheric Radial Velocity} 
\tablehead{Star & Spectral & Instrument\tablenotemark{a} & $P_{\rm rot}$ &
v(Si~IV) & v(C~IV) 
& v(Cl I) & \multicolumn{2}{c}{v(Si~IV)--v(Cl~I)} & v(C~IV)--v(Cl~I)\\
 & Type & & (days) & & & & COS & STIS & STIS}

\startdata
HII314         & G1 V    & COS    & 1.47   & --1.69 &        &--23.95 &+21.70 &
+16.25  & \\
EK Dra         & G1.5 V  & COS    & 2.61   & --1.46 &        &--17.44&+15.98 &
+10.53   & \\
$\zeta$ Dor    & F9 V    & STIS &$\leq4.0$ &  +6.54 & +8.01  &--1.67 &       &
+8.21 & +9.68\\
$\pi^1$ UMa    & G1.5 V  & COS    & 4.89   & +22.91 &        &+12.43 &+10.48  &
+5.03  & \\
$\chi^1$ Ori   & G1 V    & COS    & 5.10   & +2.20  &        &--9.31 &+11.51 &
+6.06  & \\
$\chi^1$ Ori   & G1 V    & STIS   & 5.10   &--9.45  &--8.43  &--15.51&       &
+6.06 & +7.08\\
HD 165185      & G1 V    & STIS   & 5.90   & +23.18 & +25.28 &+14.38 &       &
+8.80 & +10.90\\
HD 59967       & G3 V    & STIS   & 6.14   & +14.05 & +14.49 &+5.52  &       &
+8.53 & +8.97\\
HD 25825       & G0 V    & COS&$\approx6.5$& +65.72 &        &+52.42 &+13.30 &
+7.85  & \\
HD 97334       & G0 V    & STIS   & 7.60   & +2.32  & +2.64  &--4.48 &       &
+6.80 & +7.12\\
$\kappa^1$ Cet & G5 V    & STIS   & 8.77   & +24.11 & +25.00 &+17.08 &       &
+7.03 & +7.92\\
HD 209458      & G0 V    & COS    & 11.4   & --18.75&        &--30.90&+12.15 &
+6.70  & \\
HD 73350       & G0      & STIS   & 12.3   & +38.96 & +39.41 &+34.64 &       &
+4.32 & +4.77\\
Quiet Sun      & G2 V    & SUMER  & 25.4   &        &        &       &       &
$\approx 2.3$  & $\approx 2.6$ \\
~~~Disk Center &         & HRTS   &        &        &        &       &       &
+5.4  & +6.2 \\
Active Sun     & G2 V    & SUMER  & 25.4   &        &        &       &       &
$\approx 4.6$ & $\approx 5.4$\\
~~~Disk Center &         & HRTS   &        &        &        &       &       &
+11.0 & +13.0 \\
$\alpha$ Cen A & G2 V    & STIS   & 28.0   & --17.51& --17.56&--22.17&       &
+4.7 & +4.6\\ 

\enddata
\tablenotetext{a}{STIS velocities were measured from the StarCAT database.
COS velocity differences (v(Si IV)--v(Cl I)) are corrected 
to the STIS scale (in STARCat) by subtracting 5.45 \kms\ based on the 
comparison of $\chi^1$ Ori COS and STIS spectra. We use the 
centroid velocities of single Gaussian fits to the emission lines.}
\end{deluxetable}

\clearpage

\begin{deluxetable}{lccccccccc}
\tablewidth{0pt}
\tabletypesize{\scriptsize}
\tablenum{A1}
\rotate
\tablecaption{Emission Line Profile Parameters\tablenotemark{a} for HII314 
(COS)}
\tablehead{Ion\tablenotemark{b} & $\lambda_{lab}$ & $\lambda_{n}$ & 
$\Delta v_n$ & Flux$_n$ & FWHM$_n$ & $\lambda_{b}$ & $\Delta v_b$ & 
Flux$_b$ & FWHM$_b$\\ 
  & (\AA) & (\AA) & (km s$^{-1}$) & ($10^{-16}$) & (km s$^{-1}$) & (\AA) & 
(km s$^{-1}$) & ($10^{-16}$) & (km s$^{-1}$)}

\startdata


Fe XXI &1354.080 & $1354.0268\pm0.0073$ &$-6.86\pm1.61$ & 
$4.25\pm0.70$ & $138.6\pm10.8$ &        &      &    &\\

Fe XII&1242.000  &       &      &$<0.32$& (45.0) &       &        &      &\\
Fe XII&1349.36   &       &      &$<0.14$& (45.0) &       &        &      &\\

O V   &1218.344  & $1218.3437\pm0.0122$ & $-0.07\pm3.01$ & 
$2.31\pm0.33$ & $83.7\pm9.4$ &        &      &    &\\
O V   &1371.292  & $1371.2608\pm0.0144$ & $-6.82\pm3.16$ & 
$1.06\pm0.33$ & $46.7\pm8.9$ &        &      &    &\\

N V   &1238.821  & $1238.7850\pm0.0082$ & $-8.71\pm1.98$ & 
$17.13\pm0.99$ &  $118.3\pm4.3$ &       &      &    & \\


N V   &1242.804  & $1242.7803\pm0.0130$ & $-5.71\pm3.13$  & 
$5.62\pm0.68$& $87.7\pm6.8$ &        &      &    & \\

Si IV*&1393.755  & $1393.7440\pm0.0072$ & $-2.36\pm1.55$ & 
$27.07\pm1.94$ &  $83.9\pm4.5$ & $ 1393.6162\pm0.0317$ & 
$-29.85\pm6.82$ & $14.92\pm2.87$ & $270.5\pm35.1$\\

Si IV &1393.755  & $1393.7424\pm0.0065$ & $-2.71\pm1.40$ &
$32.05\pm1.67$&  $94.2\pm3.4$ &        &      &    &\\ 

Si IV &1402.770  & $1402.7735\pm0.0042$ & $+0.75\pm0.90$ & 
$13.91\pm0.52$ &  $93.7\pm2.6$ &        &      &    &\\



Si III*&1206.50  & $1206.4711\pm0.0080$ & $-7.18\pm2.00$ & $27.09\pm11.04$ &
$97.2\pm14.2$ & $1206.45.63\pm0.0184$ & $-10.86\pm4.58$ & 
$20.92\pm18.55$ &  $171.6\pm53.2$ \\
Si III&1206.50   & $1206.4693\pm0.0057$ & $-7.63\pm1.42$ & $49.62\pm2.10$ &
$130.7\pm3.9$ &         &      &    &\\



C II  &1335.7077 &$1335.6583\pm0.0060$ & $-11.09\pm1.35$ & $42.64\pm1.93$ &
$103.8\pm3.2$   &        &      &    &\\


O I   &1304.8576 & $1304.7382\pm0.0073$ & $-27.43\pm1.67$ & $6.39\pm1.08$ &  
$42.7\pm4.8$ &        &      &    &\\
O I   &1306.0286 & $1305.9190\pm0.0076$ & $-25.16\pm1.75$ & $5.95\pm0.98$ &  
$42.0\pm4.3$ &        &      &    &\\


Cl I  &1351.6561 & $1351.5507\pm0.0071$ & $-23.39\pm1.57$ & 
$1.93\pm0.23$ & $46.4\pm3.2$ &        &      &    &\\

\cline{1-10}

C IV*  &1548.195 & $1548.3087\pm0.0104$ & $+22.01\pm2.02$ & $49.02\pm6.92$ &
$61.9\pm 6.9$    & $1548.3772\pm0.0217$ & $+35.27\pm4.19$ & $84.78\pm12.27$ &
$222.1\pm 22.2$ \\

C IV   &1548.195 & $1548.3071\pm0.0089$ & $+21.71\pm1.72$ & $95.30\pm5.74$ &
$108.9\pm4.7$ &        &      &    &\\

C IV*  &1550.770  & $1550.9379\pm0.0083$ & $+32.48\pm1.61$ & $32.90\pm2.90$ &
$75.3\pm 5.1$     & $1550.9140\pm0.0229$ & $+27.85\pm4.43$ & $33.43\pm4.18$ &
$218.8\pm 12.1$ \\

CIV    &1550.770  & $1550.9272\pm0.0106$ & $+30.39\pm2.05$ & $53.96\pm3.69$ &
$105.5\pm4.7$ &        &      &    &\\

He II  &1640.400  & $1640.5442\pm0.0069$ & $+26.37\pm1.26$ & $107.26\pm6.05$ &
$80.8\pm2.7$ \\

C I    &1656.2672 & $1656.3593\pm0.0099$ & $+16.68\pm1.79$ & $10.25\pm1.22$ &
$52.2\pm 3.6$ \\

\enddata
\tablenotetext{a}{COS G130M data are above the break line and COS G160M data
taken later are below the break line. Parameters for the narrow components
have an ``n'' subscript and parameters for the broad components have 
a ``b'' subscript.}
\tablenotetext{b}{Spectral lines with the * symbol are preferred one-Gaussian
or two-Gaussian
fits to the observed data. Single Gaussian fits are listed below the 
two-component fits.}

\end{deluxetable}

\end{document}